\documentclass[aps,twocolumn,prd,showpacs,showkeys,preprintnumbers,nofootinbib,superscriptaddress,nobibnotes,floatfix,longbibliography]{revtex4-2}

\pdfoutput=1

\usepackage{amsmath}
\usepackage{amsfonts}
\usepackage{amssymb}
\usepackage{mathrsfs}
\usepackage{graphicx}
\usepackage{subfigure} 
\usepackage{color}
\usepackage[dvipsnames]{xcolor}
\usepackage{longtable}
\usepackage{bm}
\usepackage{blindtext}
\usepackage{wasysym}
\usepackage{hyperref}
\hypersetup{colorlinks=true,allcolors=blue}
\usepackage[normalem]{ulem}
\usepackage{lineno}
\usepackage{lipsum}
\usepackage{cancel} 

\usepackage{cuted}

\begin{document}
	

\title{Deep Earth imaging through neutrino and seismic tomography}
\thanks{A shorter version of this article is published as a meeting report in the journal {\it Current Science}~\cite{Agarwalla:2026}.}

\author{Sanjib Kumar Agarwalla}
\email{sanjib@iopb.res.in}
\affiliation{Institute of Physics, Sachivalaya Marg, Sainik School Post, Bhubaneswar 751005, India}
\affiliation{Homi Bhabha National Institute, Anushakti Nagar, Mumbai 400094, India}

\author{Arjun Datta}
\email{arjundatta@iiserpune.ac.in}
\affiliation{Indian Institute of Science Education and Research, Pune 411008, India}
\author{Amol Dighe}
\affiliation{Tata Institute of Fundamental Research, Mumbai 400005, India}
\author{Vinod Kumar Gaur}
\affiliation{CSIR Fourth Paradigm Institute, Bengaluru 560037, India}
\author{Anuj Kumar Upadhyay}
\affiliation{Institute of Physics, Sachivalaya Marg, Sainik School Post, Bhubaneswar 751005, India}
\affiliation{Department of Physics, Aligarh Muslim University, Aligarh 202002, India}

\date{\today}

\begin{abstract}
This article is a report on the Deep Earth Neutrino + Seismic Imaging and TomographY (DENSITY 2026) mini-workshop, held on 23--24 February 2026 in the Department of Earth and Climate Science at the Indian Institute of Science Education and Research (IISER), Pune. The workshop was jointly organised by IISER Pune and the Institute of Physics (IOP), Bhubaneswar. Researchers from Earth Sciences and Neutrino Physics participated in the workshop to explore multipronged approaches for studying the deep interior of the Earth. Since the participants came from diverse scientific disciplines (seismology, geochemistry, mineral physics, and neutrino physics), the programme featured a series of overview talks introducing all participants to the basic concepts of each field and highlighting how these concepts may be applied to the study of the deep Earth. 
\end{abstract}


\maketitle

The central theme of the DENSITY 2026 workshop was the emerging idea that neutrinos can serve as unique probes of the deep interior of the Earth, complementing traditional geophysical techniques such as seismic tomography. The workshop provided a platform for knowledge exchange between a section of the neutrino physics and Earth science communities in India, and possible future collaborations. The programme featured a series of talks and discussions by experts in neutrino physics, geophysics, and other interdisciplinary fields. In addition, a panel discussion was held on the proposed Underground Science Facility (USF) in India, highlighting its potential to support a broad range of multidisciplinary research, including atmospheric neutrinos, dark matter, biology, and seismology.

\section{Neutrino tomography of the Earth}
\label{sec:neutrino_tomography}

The first session of DENSITY 2026 started with a pedagogical introduction to neutrinos and their role as a unique probe of the Earth's interior. Information obtained from neutrinos, via weak interactions, is both independent of and complementary to conventional geophysical probes such as seismic and gravitational measurements. Depending on their energy, neutrinos can probe the Earth in two distinct ways: through absorption and through oscillations.

At high energies ($\ge$ tera-electron-volt, TeV), neutrinos become progressively attenuated as they traverse the Earth due to their increasing interaction probability with matter~\cite{Gandhi:1995tf}. The attenuation length depends on the nucleon number density along the neutrino trajectory. For neutrino energies above roughly a few tens of TeV, the attenuation length becomes comparable to or smaller than the Earth's diameter, leading to significant neutrino absorption. Measurements of this energy- and direction-dependent attenuation can therefore be used to infer the distribution of matter inside the Earth~\cite{Donini:2018tsg,IceCube:2026cgy}.

At lower energies, of the order of a few giga-electron-volt (GeV), neutrinos undergo flavor oscillations that are strongly affected by their interactions with electrons in ambient matter. These matter effects~\cite{Wolfenstein:1977ue,Petcov:1998su,Akhmedov:1998ui} depend on the electron number density along the neutrino trajectory, which is determined by both the matter density and the chemical composition of the Earth's layers. Atmospheric neutrinos are particularly well suited for Earth tomography because they naturally span the multi-GeV energy range where matter-induced oscillation effects become significant. In addition, they cover a wide range of propagation distances, from about 15 km to the Earth's diameter ($\sim$13000 km), passing through the crust, mantle, and even the core. Consequently, precise measurements of atmospheric neutrino oscillations can provide valuable information about the structure and composition of the Earth's deep interior~\cite{Rott:2015kwa,Kumar:2021faw,Capozzi:2021hkl,Kelly:2021jfs,Upadhyay:2021kzf,Denton:2021rgt,Upadhyay:2022jfd,Maderer:2022toi,DOlivoSaez:2022vdl,Raikwal:2023jkf,Jesus-Valls:2024tgd,Upadhyay:2024gra,Chattopadhyay:2025ulr,IceCube:2026ggn}.

Other presentations in this session highlighted recent studies using atmospheric neutrinos that have demonstrated sensitivity to several important features of the Earth's interior, including the presence of the core~\cite{Kumar:2021faw}, the location of the core-mantle boundary (CMB)~\cite{Upadhyay:2022jfd}, the densities of different internal layers correlated by the constraints of total mass and moment of inertia~\cite{Upadhyay:2024gra,Chattopadhyay:2025ulr,IceCube:2026ggn}, and even the composition of the outer core~\cite{Maderer:2022toi}. Preliminary sensitivity studies with the IceCube Upgrade detector at the South Pole indicate that, using about three years of data, neutrino oscillation measurements could constrain the density jump across the CMB to within approximately 15\%~\cite{IceCube:2026ggn}. This is particularly significant because the precise magnitude of this density discontinuity remains an open question in seismology and is not yet accurately determined through conventional geophysical measurements.

Although current neutrino tomography results are less precise than those obtained from conventional seismological methods, they provide an important proof-of-concept that neutrinos can serve as unique messengers for probing the Earth's deep interior. At present, neutrino-based Earth tomography is primarily limited by low event statistics and the need for detectors with high precision in reconstructing neutrino energy and direction. In addition, the capability to distinguish neutrinos from antineutrinos is crucial for accurately measuring matter-induced oscillation effects and for improving sensitivity to the Earth's internal density profile.

\section{Seismology and geophysical inverse theory}
\label{sec:seismology}

The second session opened with a talk that provided a pedagogical introduction to seismology and its application to studying the Earth's deep interior. Historically, seismology has been the primary tool for investigating the Earth's large-scale internal structure and has led to several major discoveries, including the existence of the dense core, the crust-mantle boundary (the Mohorovičić discontinuity, or `Moho'), the depth of CMB, and the discovery of the solid inner core.

Over the past five decades, seismological imaging capabilities have advanced significantly, evolving from one-dimensional, spherically symmetric models of the Earth’s mechanical structure (e.g., the Preliminary Reference Earth Model (PREM)~\cite{Dziewonski:1981xy}) to sophisticated three-dimensional global models (e.g., the Princeton 3-D Reference Earth Model, REM3D~\cite{Lekic:2016}). These models provide spatial distributions of elastic parameters governing seismic wave propagation, primarily the compressional-wave velocity $V_P$ and shear-wave velocity $V_S$. However, while $V_P$ and $V_S$ are well constrained by seismic observations, these observations do not uniquely determine density, which represents an additional independent parameter in the theory of elastodynamics. Consequently, despite major advances in seismic imaging, there has been relatively limited progress in constraining three-dimensional density variations inside the Earth, beyond the classical one-dimensional reference models.

Earth's density structure is most effectively studied using its normal modes, or free oscillations~\cite{Dahlen:1998,Woodhouse2015}. These low-frequency oscillations are excited by large earthquakes and cause the entire planet to `ring like a bell' at discrete eigen-frequencies determined by its internal structure. Because of their very long wavelengths, seismic normal modes sample the entire Earth, including the inner core, and are particularly sensitive to large-scale density variations, in contrast to propagating P- and S-waves, which are primarily sensitive to elastic properties. The theoretical foundation of seismic normal-mode tomography lies in the splitting of mode multiplets caused by three-dimensional heterogeneity within the Earth~\cite{Woodhouse:1978,Dahlen:1998}. This talk concluded with a discussion of how observations of normal-mode splitting can be inverted to infer the three-dimensional structure and density heterogeneity of the Earth's interior.

Inverse theory forms a cornerstone of seismic tomography and, more broadly, of geophysics, which fundamentally relies on remote observations to infer the properties of the Earth's interior. The next talk, therefore, presented an overview of geophysical inverse theory in the context of normal-mode seismology. It began by discussing the inherent challenges of inverse modelling arising from the ill-posed and nonlinear nature of the problem. In practice, deterministic solutions to nonlinear inverse problems are typically obtained by iteratively updating an initial model and searching for a local minimum in parameter space. Determining the Earth's internal density distribution is an important application of such inverse modelling techniques, where seismic normal modes excited by large earthquakes serve as the primary observational data. Conventionally, this problem is approached in two stages~\cite{giardini1987a,Giardini:1991}. First, the so-called `splitting functions', which characterise the perturbations of normal-mode frequencies due to three-dimensional heterogeneity, are estimated from observations. These splitting functions are then inverted to infer variations in the Earth's mechanical structure, particularly the shear-wave velocity and density distributions.

\section{Mineral physics and geochemistry}
\label{sec:mineral_physics}

A third session was devoted to other branches of Earth science relevant to the study of the deep Earth, particularly mineral physics and geochemistry. Mineral physics seeks to understand the behaviour of Earth materials, especially rock-forming minerals, under the extreme pressure and temperature conditions prevailing in the Earth's interior. In this context, quantum-chemical density functional theory (DFT) has emerged as a powerful computational tool. By using electronic density rather than complex many-body wave functions, DFT enables accurate predictions of the ground state properties of materials. Over the past three decades, DFT has been extensively employed to investigate the structural, electronic, magnetic, thermoelastic, and thermodynamic properties of minerals, as well as their phase relations and defect behaviour. These studies have significantly advanced our understanding of the Earth's deep interior, where direct laboratory experiments are often extremely challenging.

\begin{figure*}
	\includegraphics[width=0.9\linewidth]{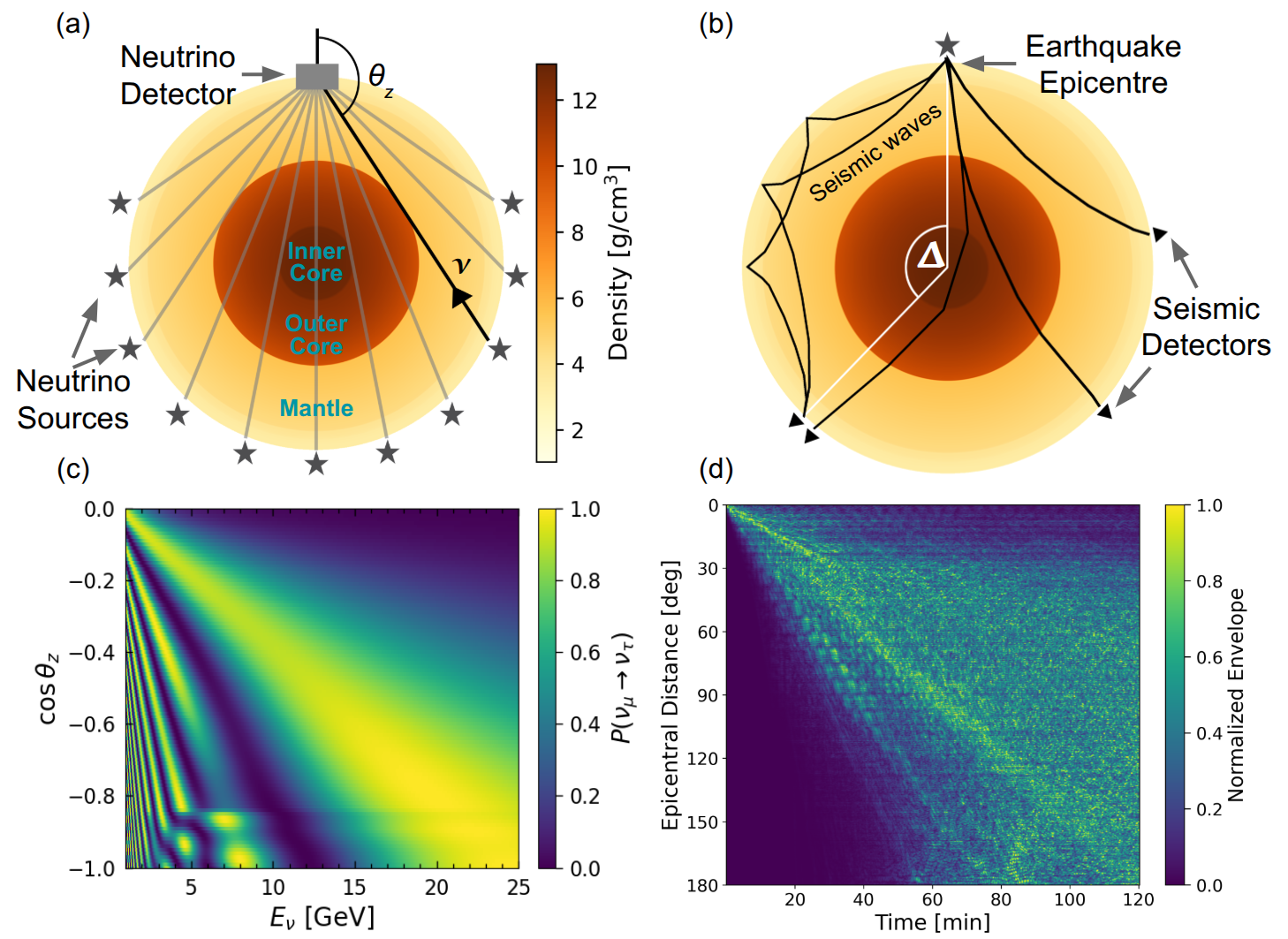}
	\caption{An example of the data-taking setups and data visualization techniques used in neutrino oscillation tomography and seismology. In (d), the normalization is carried out separately for each epicentral distance.}
	\label{fig:nu_seismic_plot}
\end{figure*}

The session concluded with a discussion on geochemistry as a complementary probe of the Earth's inaccessible core and mantle. Since direct sampling of these regions is impossible, geochemists rely on mantle-derived magmas, mantle xenoliths, and isotope geochemistry to infer their composition and evolution. Long-lived radioactive isotopes and noble gas isotopes have revealed that plate tectonics has generated chemically distinct mantle reservoirs over billions of years, including a relatively primitive lower mantle and a more processed upper mantle. Furthermore, ocean-island and continental volcanism preserve signatures of deep-mantle heterogeneity and possible interactions at the core-mantle boundary, providing evidence for a dynamically evolving and compositionally stratified Earth interior.

\section{Summary and conclusions}
\label{sec:conclusion}

The deep interior of the Earth remains largely inaccessible to direct exploration by mankind. Despite these challenges, developing a thorough understanding of the Earth's internal structure and behavior is essential for accurately interpreting the processes that occur at the surface. As discussed, neutrino oscillation tomography and seismology use different techniques to probe the deep interior of the Earth: the former relies on the oscillations or absorption of neutrinos traversing the Earth, while the latter is based on the propagation of seismic waves inside the Earth. Figures~\ref{fig:nu_seismic_plot}(a) and~\ref{fig:nu_seismic_plot}(b) illustrate the data-taking configurations for the neutrino tomography and seismology, respectively, while Figures~\ref{fig:nu_seismic_plot}(c) and~\ref{fig:nu_seismic_plot}(d) present the corresponding data visualization techniques.

As shown in Figure~\ref{fig:nu_seismic_plot}(a), neutrino tomography uses an isotropic flux of neutrinos produced in the Earth's atmosphere (multiple sources as shown by stars) and observed by a single detector (rectangular box). The black line indicates the neutrino path corresponding to the incident zenith angle $\theta_z$. This angle determines the neutrino path length from its production in the Earth's atmosphere to the detector, and hence, provides information about the electron density profile encountered along its path. On the other hand, as illustrated in Figure~\ref{fig:nu_seismic_plot}(b), seismic waves originated from a single earthquake source (star) are recorded by multiple seismic stations located at different places (triangles). The angular distance between the source and receiver, known as the epicentral distance, is denoted by $\Delta$.

As far as data visualization techniques are concerned, neutrino oscillation probabilities can be visualized in a 2D plot as functions of neutrino energy ($E_\nu$) and the incoming direction angle ($\theta_z$). Figure~\ref{fig:nu_seismic_plot}(c) shows the oscillation probability $P(\nu_\mu\rightarrow\nu_\tau)$ of a muon neutrino produced in the Earth's atmosphere being detected as a tau neutrino after propagating through different regions of the Earth, assuming the PREM density model. On the other hand, Figure~\ref{fig:nu_seismic_plot}(d) shows energies of seismic waves reaching the seismic detectors as a function of time and epicentral distance. Note that the apparent visual similarity between the Figures~\ref{fig:nu_seismic_plot}(c) and~\ref{fig:nu_seismic_plot}(d) is just a coincidence and does not reflect a fundamental relationship between them.

The workshop concluded that the methods used to probe Earth's internal structure via neutrinos and seismic waves are effectively independent and, in a sense, complementary, making them powerful and distinct approaches. The future goal of such interdisciplinary meetings is to bring together leading experts from neutrino physics and the Earth science community to explore new ways of combining their methodologies and data toward the common objective of achieving multi-messenger tomography of the Earth.

The workshop ended with a discussion on the need for an `underground science facility' in India. Both neutrino and seismic probes are strong candidates for underground laboratories: neutrino experiments need such environments to shield detectors from cosmic-ray muon backgrounds, while seismological measurements benefit from reduced surface noise caused by wind, weather, and human activity.

\begin{acknowledgements}
We warmly thank Visrutha C., R. Chakrabarti, S. Chatterjee, S. Chattopadhyay, B. Dasgupta, R. Dehiya, J. Krishnamoorthi, and A. Kumar for their valuable contributions to the workshop. We especially thank Visrutha C. for preparing one of the plots used in this manuscript. We also acknowledge the support of the Department of Atomic Energy (DAE), the Department of Science and Technology (DST), and the Anusandhan National Research Foundation (ANRF), Government of India.
\end{acknowledgements}

\bibliography{References.bib}

\begin{thebibliography}{28}%
\makeatletter
\providecommand \@ifxundefined [1]{%
 \@ifx{#1\undefined}
}%
\providecommand \@ifnum [1]{%
 \ifnum #1\expandafter \@firstoftwo
 \else \expandafter \@secondoftwo
 \fi
}%
\providecommand \@ifx [1]{%
 \ifx #1\expandafter \@firstoftwo
 \else \expandafter \@secondoftwo
 \fi
}%
\providecommand \natexlab [1]{#1}%
\providecommand \enquote  [1]{``#1''}%
\providecommand \bibnamefont  [1]{#1}%
\providecommand \bibfnamefont [1]{#1}%
\providecommand \citenamefont [1]{#1}%
\providecommand \href@noop [0]{\@secondoftwo}%
\providecommand \href [0]{\begingroup \@sanitize@url \@href}%
\providecommand \@href[1]{\@@startlink{#1}\@@href}%
\providecommand \@@href[1]{\endgroup#1\@@endlink}%
\providecommand \@sanitize@url [0]{\catcode `\\12\catcode `\$12\catcode
  `\&12\catcode `\#12\catcode `\^12\catcode `\_12\catcode `\%12\relax}%
\providecommand \@@startlink[1]{}%
\providecommand \@@endlink[0]{}%
\providecommand \url  [0]{\begingroup\@sanitize@url \@url }%
\providecommand \@url [1]{\endgroup\@href {#1}{\urlprefix }}%
\providecommand \urlprefix  [0]{URL }%
\providecommand \Eprint [0]{\href }%
\providecommand \doibase [0]{https://doi.org/}%
\providecommand \selectlanguage [0]{\@gobble}%
\providecommand \bibinfo  [0]{\@secondoftwo}%
\providecommand \bibfield  [0]{\@secondoftwo}%
\providecommand \translation [1]{[#1]}%
\providecommand \BibitemOpen [0]{}%
\providecommand \bibitemStop [0]{}%
\providecommand \bibitemNoStop [0]{.\EOS\space}%
\providecommand \EOS [0]{\spacefactor3000\relax}%
\providecommand \BibitemShut  [1]{\csname bibitem#1\endcsname}%
\let\auto@bib@innerbib\@empty
\bibitem [{\citenamefont {Agarwalla}\ \emph {et~al.}()\citenamefont
  {Agarwalla}, \citenamefont {Datta}, \citenamefont {Dighe}, \citenamefont
  {Gaur},\ and\ \citenamefont {Upadhyay}}]{Agarwalla:2026}%
  \BibitemOpen
  \bibfield  {author} {\bibinfo {author} {\bibfnamefont {S.~K.}\ \bibnamefont
  {Agarwalla}}, \bibinfo {author} {\bibfnamefont {A.}~\bibnamefont {Datta}},
  \bibinfo {author} {\bibfnamefont {A.}~\bibnamefont {Dighe}}, \bibinfo
  {author} {\bibfnamefont {V.~K.}\ \bibnamefont {Gaur}},\ and\ \bibinfo
  {author} {\bibfnamefont {A.~K.}\ \bibnamefont {Upadhyay}},\ }\bibfield
  {title} {\bibinfo {title} {{Deep Earth imaging through neutrino and seismic
  tomography}},\ }\href {https://currentscience.ac.in/Volumes/131/01/0012.pdf}
  {\bibfield  {journal} {\bibinfo  {journal} {Current Science}\ }\textbf
  {\bibinfo {volume} {131}}}\BibitemShut {NoStop}%
\bibitem [{\citenamefont {Gandhi}\ \emph {et~al.}(1996)\citenamefont {Gandhi},
  \citenamefont {Quigg}, \citenamefont {Reno},\ and\ \citenamefont
  {Sarcevic}}]{Gandhi:1995tf}%
  \BibitemOpen
  \bibfield  {author} {\bibinfo {author} {\bibfnamefont {R.}~\bibnamefont
  {Gandhi}}, \bibinfo {author} {\bibfnamefont {C.}~\bibnamefont {Quigg}},
  \bibinfo {author} {\bibfnamefont {M.~H.}\ \bibnamefont {Reno}},\ and\
  \bibinfo {author} {\bibfnamefont {I.}~\bibnamefont {Sarcevic}},\ }\bibfield
  {title} {\bibinfo {title} {{Ultrahigh-energy neutrino interactions}},\ }\href
  {https://doi.org/10.1016/0927-6505(96)00008-4} {\bibfield  {journal}
  {\bibinfo  {journal} {Astropart. Phys.}\ }\textbf {\bibinfo {volume} {5}},\
  \bibinfo {pages} {81} (\bibinfo {year} {1996})},\ \Eprint
  {https://arxiv.org/abs/hep-ph/9512364} {arXiv:hep-ph/9512364} \BibitemShut
  {NoStop}%
\bibitem [{\citenamefont {Donini}\ \emph {et~al.}(2019)\citenamefont {Donini},
  \citenamefont {Palomares-Ruiz},\ and\ \citenamefont
  {Salvado}}]{Donini:2018tsg}%
  \BibitemOpen
  \bibfield  {author} {\bibinfo {author} {\bibfnamefont {A.}~\bibnamefont
  {Donini}}, \bibinfo {author} {\bibfnamefont {S.}~\bibnamefont
  {Palomares-Ruiz}},\ and\ \bibinfo {author} {\bibfnamefont {J.}~\bibnamefont
  {Salvado}},\ }\bibfield  {title} {\bibinfo {title} {{Neutrino tomography of
  Earth}},\ }\href {https://doi.org/10.1038/s41567-018-0319-1} {\bibfield
  {journal} {\bibinfo  {journal} {Nature Phys.}\ }\textbf {\bibinfo {volume}
  {15}},\ \bibinfo {pages} {37} (\bibinfo {year} {2019})},\ \Eprint
  {https://arxiv.org/abs/1803.05901} {arXiv:1803.05901 [hep-ph]} \BibitemShut
  {NoStop}%
\bibitem [{\citenamefont {Abbasi}\ \emph
  {et~al.}(2026{\natexlab{a}})\citenamefont {Abbasi} \emph
  {et~al.}}]{IceCube:2026cgy}%
  \BibitemOpen
  \bibfield  {author} {\bibinfo {author} {\bibfnamefont {R.}~\bibnamefont
  {Abbasi}} \emph {et~al.} (\bibinfo {collaboration} {IceCube}),\ }\bibfield
  {title} {\bibinfo {title} {{High-Energy Neutrino Tomography of the Earth's
  Interior with IceCube}},\ }\href@noop {} {\  (\bibinfo {year}
  {2026}{\natexlab{a}})},\ \Eprint {https://arxiv.org/abs/2607.02644}
  {arXiv:2607.02644 [astro-ph.HE]} \BibitemShut {NoStop}%
\bibitem [{\citenamefont {Wolfenstein}(1978)}]{Wolfenstein:1977ue}%
  \BibitemOpen
  \bibfield  {author} {\bibinfo {author} {\bibfnamefont {L.}~\bibnamefont
  {Wolfenstein}},\ }\bibfield  {title} {\bibinfo {title} {{Neutrino
  Oscillations in Matter}},\ }\href {https://doi.org/10.1103/PhysRevD.17.2369}
  {\bibfield  {journal} {\bibinfo  {journal} {Phys. Rev. D}\ }\textbf {\bibinfo
  {volume} {17}},\ \bibinfo {pages} {2369} (\bibinfo {year}
  {1978})}\BibitemShut {NoStop}%
\bibitem [{\citenamefont {Petcov}(1998)}]{Petcov:1998su}%
  \BibitemOpen
  \bibfield  {author} {\bibinfo {author} {\bibfnamefont {S.~T.}\ \bibnamefont
  {Petcov}},\ }\bibfield  {title} {\bibinfo {title} {{Diffractive - like (or
  parametric resonance - like?) enhancement of the earth (day - night) effect
  for solar neutrinos crossing the earth core}},\ }\href
  {https://doi.org/10.1016/S0370-2693(98)00742-4} {\bibfield  {journal}
  {\bibinfo  {journal} {Phys. Lett. B}\ }\textbf {\bibinfo {volume} {434}},\
  \bibinfo {pages} {321} (\bibinfo {year} {1998})},\ \Eprint
  {https://arxiv.org/abs/hep-ph/9805262} {arXiv:hep-ph/9805262} \BibitemShut
  {NoStop}%
\bibitem [{\citenamefont {Akhmedov}(1999)}]{Akhmedov:1998ui}%
  \BibitemOpen
  \bibfield  {author} {\bibinfo {author} {\bibfnamefont {E.~K.}\ \bibnamefont
  {Akhmedov}},\ }\bibfield  {title} {\bibinfo {title} {{Parametric resonance of
  neutrino oscillations and passage of solar and atmospheric neutrinos through
  the earth}},\ }\href {https://doi.org/10.1016/S0550-3213(98)00723-8}
  {\bibfield  {journal} {\bibinfo  {journal} {Nucl. Phys. B}\ }\textbf
  {\bibinfo {volume} {538}},\ \bibinfo {pages} {25} (\bibinfo {year} {1999})},\
  \Eprint {https://arxiv.org/abs/hep-ph/9805272} {arXiv:hep-ph/9805272}
  \BibitemShut {NoStop}%
\bibitem [{\citenamefont {Rott}\ \emph {et~al.}(2015)\citenamefont {Rott},
  \citenamefont {Taketa},\ and\ \citenamefont {Bose}}]{Rott:2015kwa}%
  \BibitemOpen
  \bibfield  {author} {\bibinfo {author} {\bibfnamefont {C.}~\bibnamefont
  {Rott}}, \bibinfo {author} {\bibfnamefont {A.}~\bibnamefont {Taketa}},\ and\
  \bibinfo {author} {\bibfnamefont {D.}~\bibnamefont {Bose}},\ }\bibfield
  {title} {\bibinfo {title} {{Spectrometry of the Earth using Neutrino
  Oscillations}},\ }\href {https://doi.org/10.1038/srep15225} {\bibfield
  {journal} {\bibinfo  {journal} {Sci. Rep.}\ }\textbf {\bibinfo {volume}
  {5}},\ \bibinfo {pages} {15225} (\bibinfo {year} {2015})},\ \Eprint
  {https://arxiv.org/abs/1502.04930} {arXiv:1502.04930 [physics.geo-ph]}
  \BibitemShut {NoStop}%
\bibitem [{\citenamefont {Kumar}\ and\ \citenamefont
  {Agarwalla}(2021)}]{Kumar:2021faw}%
  \BibitemOpen
  \bibfield  {author} {\bibinfo {author} {\bibfnamefont {A.}~\bibnamefont
  {Kumar}}\ and\ \bibinfo {author} {\bibfnamefont {S.~K.}\ \bibnamefont
  {Agarwalla}},\ }\bibfield  {title} {\bibinfo {title} {{Validating the
  Earth{\textquoteright}s core using atmospheric neutrinos with ICAL at INO}},\
  }\href {https://doi.org/10.1007/JHEP08(2021)139} {\bibfield  {journal}
  {\bibinfo  {journal} {JHEP}\ }\textbf {\bibinfo {volume} {08}},\ \bibinfo
  {pages} {139}},\ \Eprint {https://arxiv.org/abs/2104.11740} {arXiv:2104.11740
  [hep-ph]} \BibitemShut {NoStop}%
\bibitem [{\citenamefont {Capozzi}\ and\ \citenamefont
  {Petcov}(2022)}]{Capozzi:2021hkl}%
  \BibitemOpen
  \bibfield  {author} {\bibinfo {author} {\bibfnamefont {F.}~\bibnamefont
  {Capozzi}}\ and\ \bibinfo {author} {\bibfnamefont {S.~T.}\ \bibnamefont
  {Petcov}},\ }\bibfield  {title} {\bibinfo {title} {{Neutrino tomography of
  the Earth with ORCA detector}},\ }\href
  {https://doi.org/10.1140/epjc/s10052-022-10399-6} {\bibfield  {journal}
  {\bibinfo  {journal} {Eur. Phys. J. C}\ }\textbf {\bibinfo {volume} {82}},\
  \bibinfo {pages} {461} (\bibinfo {year} {2022})},\ \Eprint
  {https://arxiv.org/abs/2111.13048} {arXiv:2111.13048 [hep-ph]} \BibitemShut
  {NoStop}%
\bibitem [{\citenamefont {Kelly}\ \emph {et~al.}(2022)\citenamefont {Kelly},
  \citenamefont {Machado}, \citenamefont {Martinez-Soler},\ and\ \citenamefont
  {Perez-Gonzalez}}]{Kelly:2021jfs}%
  \BibitemOpen
  \bibfield  {author} {\bibinfo {author} {\bibfnamefont {K.~J.}\ \bibnamefont
  {Kelly}}, \bibinfo {author} {\bibfnamefont {P.~A.~N.}\ \bibnamefont
  {Machado}}, \bibinfo {author} {\bibfnamefont {I.}~\bibnamefont
  {Martinez-Soler}},\ and\ \bibinfo {author} {\bibfnamefont {Y.~F.}\
  \bibnamefont {Perez-Gonzalez}},\ }\bibfield  {title} {\bibinfo {title} {{DUNE
  atmospheric neutrinos: Earth tomography}},\ }\href
  {https://doi.org/10.1007/JHEP05(2022)187} {\bibfield  {journal} {\bibinfo
  {journal} {JHEP}\ }\textbf {\bibinfo {volume} {05}},\ \bibinfo {pages}
  {187}},\ \Eprint {https://arxiv.org/abs/2110.00003} {arXiv:2110.00003
  [hep-ph]} \BibitemShut {NoStop}%
\bibitem [{\citenamefont {Upadhyay}\ \emph
  {et~al.}(2023{\natexlab{a}})\citenamefont {Upadhyay}, \citenamefont {Kumar},
  \citenamefont {Agarwalla},\ and\ \citenamefont {Dighe}}]{Upadhyay:2021kzf}%
  \BibitemOpen
  \bibfield  {author} {\bibinfo {author} {\bibfnamefont {A.~K.}\ \bibnamefont
  {Upadhyay}}, \bibinfo {author} {\bibfnamefont {A.}~\bibnamefont {Kumar}},
  \bibinfo {author} {\bibfnamefont {S.~K.}\ \bibnamefont {Agarwalla}},\ and\
  \bibinfo {author} {\bibfnamefont {A.}~\bibnamefont {Dighe}},\ }\bibfield
  {title} {\bibinfo {title} {{Probing dark matter inside Earth using
  atmospheric neutrino oscillations at INO-ICAL}},\ }\href
  {https://doi.org/10.1103/PhysRevD.107.115030} {\bibfield  {journal} {\bibinfo
   {journal} {Phys. Rev. D}\ }\textbf {\bibinfo {volume} {107}},\ \bibinfo
  {pages} {115030} (\bibinfo {year} {2023}{\natexlab{a}})},\ \Eprint
  {https://arxiv.org/abs/2112.14201} {arXiv:2112.14201 [hep-ph]} \BibitemShut
  {NoStop}%
\bibitem [{\citenamefont {Denton}\ and\ \citenamefont
  {Pestes}(2021)}]{Denton:2021rgt}%
  \BibitemOpen
  \bibfield  {author} {\bibinfo {author} {\bibfnamefont {P.~B.}\ \bibnamefont
  {Denton}}\ and\ \bibinfo {author} {\bibfnamefont {R.}~\bibnamefont
  {Pestes}},\ }\bibfield  {title} {\bibinfo {title} {{Neutrino oscillations
  through the Earth{\textquoteright}s core}},\ }\href
  {https://doi.org/10.1103/PhysRevD.104.113007} {\bibfield  {journal} {\bibinfo
   {journal} {Phys. Rev. D}\ }\textbf {\bibinfo {volume} {104}},\ \bibinfo
  {pages} {113007} (\bibinfo {year} {2021})},\ \Eprint
  {https://arxiv.org/abs/2110.01148} {arXiv:2110.01148 [hep-ph]} \BibitemShut
  {NoStop}%
\bibitem [{\citenamefont {Upadhyay}\ \emph
  {et~al.}(2023{\natexlab{b}})\citenamefont {Upadhyay}, \citenamefont {Kumar},
  \citenamefont {Agarwalla},\ and\ \citenamefont {Dighe}}]{Upadhyay:2022jfd}%
  \BibitemOpen
  \bibfield  {author} {\bibinfo {author} {\bibfnamefont {A.~K.}\ \bibnamefont
  {Upadhyay}}, \bibinfo {author} {\bibfnamefont {A.}~\bibnamefont {Kumar}},
  \bibinfo {author} {\bibfnamefont {S.~K.}\ \bibnamefont {Agarwalla}},\ and\
  \bibinfo {author} {\bibfnamefont {A.}~\bibnamefont {Dighe}},\ }\bibfield
  {title} {\bibinfo {title} {{Locating the core-mantle boundary using
  oscillations of atmospheric neutrinos}},\ }\href
  {https://doi.org/10.1007/JHEP04(2023)068} {\bibfield  {journal} {\bibinfo
  {journal} {JHEP}\ }\textbf {\bibinfo {volume} {04}},\ \bibinfo {pages}
  {068}},\ \Eprint {https://arxiv.org/abs/2211.08688} {arXiv:2211.08688
  [hep-ph]} \BibitemShut {NoStop}%
\bibitem [{\citenamefont {Maderer}\ \emph {et~al.}(2023)\citenamefont
  {Maderer}, \citenamefont {Kaminski}, \citenamefont {Coelho}, \citenamefont
  {Bourret},\ and\ \citenamefont {Van~Elewyck}}]{Maderer:2022toi}%
  \BibitemOpen
  \bibfield  {author} {\bibinfo {author} {\bibfnamefont {L.}~\bibnamefont
  {Maderer}}, \bibinfo {author} {\bibfnamefont {E.}~\bibnamefont {Kaminski}},
  \bibinfo {author} {\bibfnamefont {J.~A.~B.}\ \bibnamefont {Coelho}}, \bibinfo
  {author} {\bibfnamefont {S.}~\bibnamefont {Bourret}},\ and\ \bibinfo {author}
  {\bibfnamefont {V.}~\bibnamefont {Van~Elewyck}},\ }\bibfield  {title}
  {\bibinfo {title} {{Unveiling the outer core composition with neutrino
  oscillation tomography}},\ }\href
  {https://doi.org/10.3389/feart.2023.1008396} {\bibfield  {journal} {\bibinfo
  {journal} {Front. Earth Sci.}\ }\textbf {\bibinfo {volume} {11}},\ \bibinfo
  {pages} {1008396} (\bibinfo {year} {2023})},\ \Eprint
  {https://arxiv.org/abs/2208.00532} {arXiv:2208.00532 [hep-ex]} \BibitemShut
  {NoStop}%
\bibitem [{\citenamefont {D'Olivo~Saez}\ \emph {et~al.}(2022)\citenamefont
  {D'Olivo~Saez}, \citenamefont {Lara}, \citenamefont {Romero},\ and\
  \citenamefont {Sampayo}}]{DOlivoSaez:2022vdl}%
  \BibitemOpen
  \bibfield  {author} {\bibinfo {author} {\bibfnamefont {J.~C.}\ \bibnamefont
  {D'Olivo~Saez}}, \bibinfo {author} {\bibfnamefont {J.~A.~H.}\ \bibnamefont
  {Lara}}, \bibinfo {author} {\bibfnamefont {I.}~\bibnamefont {Romero}},\ and\
  \bibinfo {author} {\bibfnamefont {O.~A.}\ \bibnamefont {Sampayo}},\
  }\bibfield  {title} {\bibinfo {title} {{Oscillation tomografy study of
  Earth{\textquoteright}s composition and density with atmospheric
  neutrinos}},\ }\href {https://doi.org/10.1140/epjc/s10052-022-10563-y}
  {\bibfield  {journal} {\bibinfo  {journal} {Eur. Phys. J. C}\ }\textbf
  {\bibinfo {volume} {82}},\ \bibinfo {pages} {614} (\bibinfo {year} {2022})},\
  \Eprint {https://arxiv.org/abs/2207.11257} {arXiv:2207.11257
  [physics.geo-ph]} \BibitemShut {NoStop}%
\bibitem [{\citenamefont {Raikwal}\ and\ \citenamefont
  {Choubey}(2024)}]{Raikwal:2023jkf}%
  \BibitemOpen
  \bibfield  {author} {\bibinfo {author} {\bibfnamefont {D.}~\bibnamefont
  {Raikwal}}\ and\ \bibinfo {author} {\bibfnamefont {S.}~\bibnamefont
  {Choubey}},\ }\bibfield  {title} {\bibinfo {title} {{Earth tomography with
  the ICAL detector at INO}},\ }\href
  {https://doi.org/10.1103/PhysRevD.109.073011} {\bibfield  {journal} {\bibinfo
   {journal} {Phys. Rev. D}\ }\textbf {\bibinfo {volume} {109}},\ \bibinfo
  {pages} {073011} (\bibinfo {year} {2024})},\ \Eprint
  {https://arxiv.org/abs/2309.12573} {arXiv:2309.12573 [hep-ph]} \BibitemShut
  {NoStop}%
\bibitem [{\citenamefont {Jes{\'u}s-Valls}\ \emph {et~al.}(2025)\citenamefont
  {Jes{\'u}s-Valls}, \citenamefont {Petcov},\ and\ \citenamefont
  {Xia}}]{Jesus-Valls:2024tgd}%
  \BibitemOpen
  \bibfield  {author} {\bibinfo {author} {\bibfnamefont {C.}~\bibnamefont
  {Jes{\'u}s-Valls}}, \bibinfo {author} {\bibfnamefont {S.~T.}\ \bibnamefont
  {Petcov}},\ and\ \bibinfo {author} {\bibfnamefont {J.}~\bibnamefont {Xia}},\
  }\bibfield  {title} {\bibinfo {title} {{Neutrino oscillation tomography of
  the Earth with the Hyper-Kamiokande detector}},\ }\href
  {https://doi.org/10.1140/epjc/s10052-025-14338-z} {\bibfield  {journal}
  {\bibinfo  {journal} {Eur. Phys. J. C}\ }\textbf {\bibinfo {volume} {85}},\
  \bibinfo {pages} {703} (\bibinfo {year} {2025})},\ \Eprint
  {https://arxiv.org/abs/2411.12344} {arXiv:2411.12344 [hep-ex]} \BibitemShut
  {NoStop}%
\bibitem [{\citenamefont {Upadhyay}\ \emph {et~al.}(2026)\citenamefont
  {Upadhyay}, \citenamefont {Kumar}, \citenamefont {Agarwalla},\ and\
  \citenamefont {Dighe}}]{Upadhyay:2024gra}%
  \BibitemOpen
  \bibfield  {author} {\bibinfo {author} {\bibfnamefont {A.~K.}\ \bibnamefont
  {Upadhyay}}, \bibinfo {author} {\bibfnamefont {A.}~\bibnamefont {Kumar}},
  \bibinfo {author} {\bibfnamefont {S.~K.}\ \bibnamefont {Agarwalla}},\ and\
  \bibinfo {author} {\bibfnamefont {A.}~\bibnamefont {Dighe}},\ }\bibfield
  {title} {\bibinfo {title} {{Constraining the core radius and density jumps
  inside Earth using atmospheric neutrino oscillations}},\ }\href
  {https://doi.org/10.1007/JHEP04(2026)066} {\bibfield  {journal} {\bibinfo
  {journal} {JHEP}\ }\textbf {\bibinfo {volume} {04}},\ \bibinfo {pages}
  {066}},\ \Eprint {https://arxiv.org/abs/2405.04986} {arXiv:2405.04986
  [hep-ph]} \BibitemShut {NoStop}%
\bibitem [{\citenamefont {Chattopadhyay}\ \emph {et~al.}(2025)\citenamefont
  {Chattopadhyay}, \citenamefont {Krishnamoorthi},\ and\ \citenamefont
  {Upadhyay}}]{Chattopadhyay:2025ulr}%
  \BibitemOpen
  \bibfield  {author} {\bibinfo {author} {\bibfnamefont {S.}~\bibnamefont
  {Chattopadhyay}}, \bibinfo {author} {\bibfnamefont {J.}~\bibnamefont
  {Krishnamoorthi}},\ and\ \bibinfo {author} {\bibfnamefont {A.~K.}\
  \bibnamefont {Upadhyay}},\ }\bibfield  {title} {\bibinfo {title}
  {{Demonstrating the ability of IceCube DeepCore to probe
  Earth{\textquoteright}s interior with atmospheric neutrino oscillations}},\
  }\href {https://doi.org/10.1140/epjs/s11734-025-01741-6} {\bibfield
  {journal} {\bibinfo  {journal} {Eur. Phys. J. ST}\ }\textbf {\bibinfo
  {volume} {234}},\ \bibinfo {pages} {5055} (\bibinfo {year} {2025})},\ \Eprint
  {https://arxiv.org/abs/2502.18995} {arXiv:2502.18995 [hep-ph]} \BibitemShut
  {NoStop}%
\bibitem [{\citenamefont {Abbasi}\ \emph
  {et~al.}(2026{\natexlab{b}})\citenamefont {Abbasi} \emph
  {et~al.}}]{IceCube:2026ggn}%
  \BibitemOpen
  \bibfield  {author} {\bibinfo {author} {\bibfnamefont {R.}~\bibnamefont
  {Abbasi}} \emph {et~al.} (\bibinfo {collaboration} {IceCube}),\ }\bibfield
  {title} {\bibinfo {title} {{Estimating the sensitivity of the IceCube Upgrade
  to probe the interior of the Earth using atmospheric neutrino
  oscillations}},\ }\href@noop {} {\  (\bibinfo {year} {2026}{\natexlab{b}})},\
  \Eprint {https://arxiv.org/abs/2608.06543} {arXiv:2608.06543 [hep-ex]}
  \BibitemShut {NoStop}%
\bibitem [{\citenamefont {Dziewonski}\ and\ \citenamefont
  {Anderson}(1981)}]{Dziewonski:1981xy}%
  \BibitemOpen
  \bibfield  {author} {\bibinfo {author} {\bibfnamefont {A.~M.}\ \bibnamefont
  {Dziewonski}}\ and\ \bibinfo {author} {\bibfnamefont {D.~L.}\ \bibnamefont
  {Anderson}},\ }\bibfield  {title} {\bibinfo {title} {{Preliminary reference
  earth model}},\ }\href {https://doi.org/10.1016/0031-9201(81)90046-7}
  {\bibfield  {journal} {\bibinfo  {journal} {Phys. Earth Planet. Interiors}\
  }\textbf {\bibinfo {volume} {25}},\ \bibinfo {pages} {297} (\bibinfo {year}
  {1981})}\BibitemShut {NoStop}%
\bibitem [{\citenamefont {{Lekic}}\ \emph {et~al.}(2016)\citenamefont
  {{Lekic}}, \citenamefont {{Moulik}}, \citenamefont {{Romanowicz}},\ and\
  \citenamefont {{Dziewonski}}}]{Lekic:2016}%
  \BibitemOpen
  \bibfield  {author} {\bibinfo {author} {\bibfnamefont {V.}~\bibnamefont
  {{Lekic}}}, \bibinfo {author} {\bibfnamefont {P.}~\bibnamefont {{Moulik}}},
  \bibinfo {author} {\bibfnamefont {B.~A.}\ \bibnamefont {{Romanowicz}}},\ and\
  \bibinfo {author} {\bibfnamefont {A.~M.}\ \bibnamefont {{Dziewonski}}},\
  }\bibfield  {title} {\bibinfo {title} {{The 3D Reference Earth Model
  (REM-3D): Update and Outlook}},\ }in\ \href@noop {} {\emph {\bibinfo
  {booktitle} {AGU Fall Meeting Abstracts}}},\ \bibinfo {series} {AGU Fall
  Meeting Abstracts}, Vol.\ \bibinfo {volume} {2016}\ (\bibinfo {year} {2016})\
  pp.\ \bibinfo {pages} {DI31A--2617}\BibitemShut {NoStop}%
\bibitem [{\citenamefont {Dahlen}\ and\ \citenamefont
  {Tromp}(1998)}]{Dahlen:1998}%
  \BibitemOpen
  \bibfield  {author} {\bibinfo {author} {\bibfnamefont {F.~A.}\ \bibnamefont
  {Dahlen}}\ and\ \bibinfo {author} {\bibfnamefont {J.}~\bibnamefont {Tromp}},\
  }\href {https://doi.org/https://doi.org/10.2307/j.ctv131bvfd} {\emph
  {\bibinfo {title} {Theoretical Global Seismology}}}\ (\bibinfo  {publisher}
  {Princeton University Press},\ \bibinfo {year} {1998})\BibitemShut {NoStop}%
\bibitem [{\citenamefont {Woodhouse}\ and\ \citenamefont
  {Deuss}(2015)}]{Woodhouse2015}%
  \BibitemOpen
  \bibfield  {author} {\bibinfo {author} {\bibfnamefont {J.~H.}\ \bibnamefont
  {Woodhouse}}\ and\ \bibinfo {author} {\bibfnamefont {A.}~\bibnamefont
  {Deuss}},\ }\bibinfo {title} {Theory and observations -- earth's free
  oscillations},\ in\ \href
  {https://doi.org/10.1016/B978-0-444-53802-4.00002-6} {\emph {\bibinfo
  {booktitle} {Treatise on Geophysics, Volume 1: Deep Earth Seismology}}},\
  Vol.~\bibinfo {volume} {1}\ (\bibinfo {year} {2015})\ pp.\ \bibinfo {pages}
  {79--115},\ \bibinfo {edition} {second edition}\ ed.\BibitemShut {Stop}%
\bibitem [{\citenamefont {Woodhouse}\ and\ \citenamefont
  {Dahlen}(1978)}]{Woodhouse:1978}%
  \BibitemOpen
  \bibfield  {author} {\bibinfo {author} {\bibfnamefont {J.~H.}\ \bibnamefont
  {Woodhouse}}\ and\ \bibinfo {author} {\bibfnamefont {F.~A.}\ \bibnamefont
  {Dahlen}},\ }\bibfield  {title} {\bibinfo {title} {The effect of a general
  aspherical perturbation on the free oscillations of the earth},\ }\href
  {https://doi.org/10.1111/j.1365-246X.1978.tb03746.x} {\bibfield  {journal}
  {\bibinfo  {journal} {Geophysical Journal International}\ }\textbf {\bibinfo
  {volume} {53}},\ \bibinfo {pages} {335} (\bibinfo {year} {1978})}\BibitemShut
  {NoStop}%
\bibitem [{\citenamefont {Giardini}\ \emph {et~al.}(1987)\citenamefont
  {Giardini}, \citenamefont {Li},\ and\ \citenamefont
  {Woodhouse}}]{giardini1987a}%
  \BibitemOpen
  \bibfield  {author} {\bibinfo {author} {\bibfnamefont {D.}~\bibnamefont
  {Giardini}}, \bibinfo {author} {\bibfnamefont {X.}~\bibnamefont {Li}},\ and\
  \bibinfo {author} {\bibfnamefont {J.}~\bibnamefont {Woodhouse}},\ }\bibfield
  {title} {\bibinfo {title} {Three-dimensional structure of the earth from
  splitting in free-oscillation spectra},\ }\href@noop {} {\bibfield  {journal}
  {\bibinfo  {journal} {Nature}\ }\textbf {\bibinfo {volume} {325}},\ \bibinfo
  {pages} {405} (\bibinfo {year} {1987})}\BibitemShut {NoStop}%
\bibitem [{\citenamefont {Li}\ \emph {et~al.}(1991)\citenamefont {Li},
  \citenamefont {Giardini},\ and\ \citenamefont {Woodhouses}}]{Giardini:1991}%
  \BibitemOpen
  \bibfield  {author} {\bibinfo {author} {\bibfnamefont {X.-D.}\ \bibnamefont
  {Li}}, \bibinfo {author} {\bibfnamefont {D.}~\bibnamefont {Giardini}},\ and\
  \bibinfo {author} {\bibfnamefont {J.~H.}\ \bibnamefont {Woodhouses}},\
  }\bibfield  {title} {\bibinfo {title} {Large-scale three-dimensional
  even-degree structure of the earth from splitting of long-period normal
  modes},\ }\href {https://doi.org/https://doi.org/10.1029/90JB02009}
  {\bibfield  {journal} {\bibinfo  {journal} {Journal of Geophysical Research:
  Solid Earth}\ }\textbf {\bibinfo {volume} {96}},\ \bibinfo {pages} {551}
  (\bibinfo {year} {1991})}\BibitemShut {NoStop}%
\end{thebibliography}%

\end{document}